\documentclass[journal,twocolumn,transmag]{IEEEtran}
\usepackage{cite}
\usepackage[cmex10]{amsmath}
\usepackage{amssymb}
\usepackage{lscape}
\usepackage[font={small}]{caption}
\usepackage{caption}
\usepackage{algorithm}
\usepackage{algorithmic}
\graphicspath{{./figures/}}
\usepackage{graphicx}
\usepackage{float}
\usepackage{rotating}
\usepackage{subfig}
\usepackage{booktabs}
 \usepackage{multirow}
 \usepackage{hyperref}
 \usepackage{url} 

\begin{document}
%
\title{Design and Implementation of Performance Metrics for Evaluation of Assessments Data}


\author{\IEEEauthorblockN{Irfan~Ahmed \,
and
Arif~Bhatti
}
\IEEEauthorblockA{College of Computers and Information Technology,
Taif University, Taif 21974 KSA}}


%

\newpage

\IEEEtitleabstractindextext{%
\begin{abstract}
The objective of this paper is to design performance metrics and respective formulas to quantitatively evaluate the achievement of set objectives and expected outcomes both at the course and program levels. Evaluation is defined as one or more processes for interpreting the data acquired through the assessment processes in order to determine how well the set objectives and outcomes are being attained. Even though assessment processes for accreditation are well documented but existence of an evaluation process is assumed. This paper focuses on evaluation process to provide insights and techniques for data interpretation. It gives a complete evaluation process from the data collection through various assessment methods, performance metrics, to the presentations in the form of tables and graphs.  Authors hope that the articulated description of evaluation formulas will help convergence to high quality standard in evaluation process.
\end{abstract}

}

\newpage

\maketitle

\IEEEdisplaynontitleabstractindextext

%
\IEEEpeerreviewmaketitle

\section{Introduction}
%
%
%
%
\IEEEPARstart{A}{ny} educational program starts with a mission statement, objectives, and the program or student outcomes. Mission statement describes the broad goals of the program. Program educational objectives (PEO) are broad statements to describe what graduates are expected to attain within a few years of graduation.
Student outcomes (SO) describe what students are expected to know and be able to do by the time of graduation. SOs relate to the skills, knowledge, and behaviors that students acquire as they progress through the program \cite{ABET_abet:_2014}.

Assessment is defined as one or more processes that identify, collect, and prepare the data necessary for evaluation.  Evaluation is defined as one or more processes for interpreting the data acquired through the assessment processes in order to determine how well SOs are being attained \cite{ABET_abet:_2014}. The quality assurance of an academic program depends upon the quantitative and qualitative measurements of these PEOs and SOs.
The main component of the quality assurance system is the continuous improvement process. Faculty and students assess the achievement of SOs while advisory board members from industry and academia, program alumni, and employers are the main resource for accessing the PEO achievements. Continuous improvement requires regular documented assessment and evaluation of SOs. These evaluation are used as input to make changes in curriculum, teaching methodologies, and revision of SOs or PEOs.

There are many international academic accreditation bodies establishing standards, guidelines, and criteria to ensure a certain degree of quality assurance and progress towards continuous improvement. Accreditation Board for Engineering and Technology (ABET) \url{(http://www.abet.org)} is a US-based educational programs accreditation body. It is recognized as the worldwide leader in assuring quality and stimulating innovation in applied science, computing, engineering, and technology educations. Its responsibilities include ''organizing and carrying out a comprehensive process of accreditation of pertinent programs leading to degrees, and assisting academic institutions in planning their educational programs'' \cite{ABET_abet:_2014}. At the national level in Saudi Arabia, National Commission for Assessment and Accreditation (NCAAA) \url{(http://ncaaa.org.sa)} is responsible for the quality, effectiveness and continuing improvement in the quality of post secondary education. The presented evaluation methodologies are equally applicable to ABET and NCAAA accreditation processes.

Many authors have published their work on accreditation process, continuous improvement, assessment strategies but there is no work in the literature that focuses on the evaluation of the assessment data at course and program levels.
A complete procedure for ABET accreditation for Engineering programs at Qassim University has been presented in \cite{al-yahya_successful_2013}.
It gives the detailed implementation of the continuous improvement process, effecting major changes in the educational plan, curricular content, facilities, activities, teaching methodologies, and assessment practices. But this paper does not go into the details of evaluation process. In \cite{pierrakos_comprehensive_2013} authors present an assessment plan and continuous improvement process in preparation for ABET in department of engineering at James Madison University. They have introduced course assessment and continuous improvement (CACI) reports at course-level and student outcomes summary report (SOSR) at program-level. This paper shows some sample reports and assessment templates but does not discuss the evaluation process. A web-based tool has been introduced in \cite{reed_hierarchical_2013} for outcome-based open-ended and recursive hierarchical quantitative assessment. This quantitative assessment is used to structure outcomes and measures into a leveled hierarchy, with course outcomes at the bottom and more general objectives at the top. A general curriculum outcome (GCO) layer has been added between course's outcomes and program or student's outcomes.
In \cite{el-ariss_civil_2009} the authors describe the development of an Excel spreadsheet and the associated assessment tools for a technical design course to measure its success and ensure its continuous improvement to meet the requirements of the ABET engineering criteria.
In \cite{kalaani_continuous_2014} both direct and indirect measures are used to collect and analyze data to assess the attainments of the student outcomes. To ensure data integrity, a set of rubrics with benchmarks and performance indicators at both the program and curriculum levels are developed. Each outcome has been assessed for different levels (introductory, beginning, developing, proficient, exemplary) and from different sources.
The article \cite{sriraman_lessons_2013} presents discussions on writing learning outcomes and to assess soft skills in engineering education.
ABET accreditation preparations for construction engineering program at the American University in Cairo are described in \cite{ezeldin_international_2013}. It explains the mapping of department mission and objectives with university mission and objectives, respectively. Then the PEO to SO mapping and SO to course learning outcome (CLO) are provided. The paper \cite{nasir_towards_2011} describes the assessment techniques and the mapping of CLO to SO without the insight of evaluation process.
A case study \cite{Hughes_case_2013} describes the features that contribute to assessment quality at the programme, course and task level. This case study has a particular focus on the ‘technical’ such as task analysis and task relationship patterns.
Another case study \cite{Harris_Implementing_2010} presents a health science program reform and evaluation. It discusses potential for evaluation to establish responsive communication between students, teaching staff and programme administrators, ensuring a match between the intended, implemented and attained curriculum. None of these works however, provide the detail of evaluation metrics and their use in course and program assessment.
\begin{figure}[!htbp]
  \centering
  \includegraphics[width=3.5in]{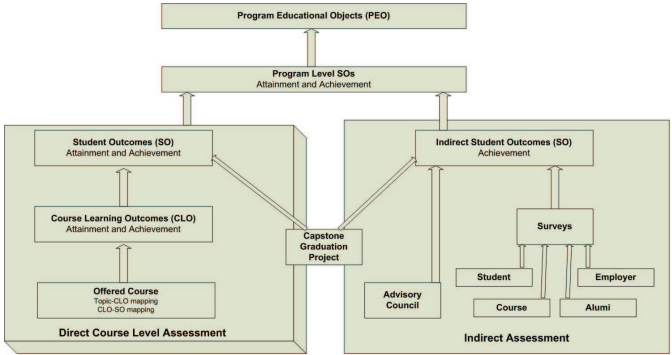}\\
  \caption{Components of direct and indirect approaches in assessment process. Direct approach depends on computation and analytical analysis of collected data}\label{AssessmentProcess}
\end{figure}

This paper presents the evaluation of assessment data obtained from direct and indirect assessments as shown in Fig. \ref{AssessmentProcess}. It gives a complete evaluation process from the data collection through various assessment methods, performance metrics, to the presentations in the form of tables and graphs.  Authors hope that the articulated description of evaluation formulas will help convergence to high quality standard in evaluation process.
The rest of the article is organized as follows: the next section describes the performance metrics formulation, course level and program level evaluation based on performance metrics are explained in section III and IV, analysis and interpretations are discussed in section V and conclusions are drawn in section VI.

\section{Performance Metrics Formulation}
Let $A_{i,j,m}$ be the marks obtained by student $m$ in question $j$ of assessment $i$ (homework, assignment, quiz, midterm, or final etc). Here $i$ can take the values, $i=1,2,...,I$, $j$  can take the values $j_{i}=1,2,...,J_{i}$, and $m$ can take the values $m=1,2,...,M$. $I,J_{i},M$ are the total number of assessments, questions in assessment $i$, and the students, respectively.

For quantitative analysis, question is a basic unit of computation for assessment. Average score of question $j$ in assessment $i$ that has $M$ students is given by
\begin{equation}\label{average}
  B_{i,j}=\frac{1}{M}\sum_{m=1}^{M}A_{i,j,m}
\end{equation}
$B_{i,j}$ can be written in vector form as
\begin{equation}\label{vectorB}
  \mathbf{\tilde{B}}=
  \left[
                    \begin{array}{ccccc}
                       \mathbf{B}_{1} & \mathbf{B}_{2} & . & . & \mathbf{B}_{I}
                     \end{array}
  \right]_{1\times L}
\end{equation}
Passing threshold (PT) could be absolute, relative or composite \cite{reed_hierarchical_2013}, such that the PT of question $j$ in assessment $i$ is given by one of the following:
\begin{equation}\label{passingThreshold1}
  PT_{i,j}=\alpha \times Q_{i,j}^{tot}
\end{equation}
\begin{equation}\label{passingThreshold2}
  PT_{i,j}=B_{i,j}
\end{equation}
\begin{equation}\label{passingThreshold}
  PT_{i,j}=\min \{B_{i,j},\alpha \times Q_{i,j}^{tot}\}
\end{equation}
where $Q_{i,j}^{tot}$ is the maximum or total marks of question $j$ in assessment $i$ and $0< \alpha <1$.
The maximum, minimum, standard deviation, and \emph{x-th} percentile of question $j$ in assessment $i$ are calculated as
\begin{eqnarray}
\nonumber  A_{i,j,max} &=& \max_{\substack{m}} \mathbf{A}_{i,j,:} \\
\nonumber  A_{i,j,min} &=& \min_{\substack{m}} \mathbf{A}_{i,j,:} \\
\nonumber  A_{i,j,std} &=& stdev \mathbf{A}_{i,j,m}\\
\nonumber  A_{i,j,per} &=& percentile(\mathbf{A}_{i,j,:},x)
\end{eqnarray}

Course learning outcomes describe what students are expected to learn in a course. A mapping between CLO and assessment questions is required to compute the attainment of the course CLOs. If a course covers $N$ number of CLOs then $n^{th}$ CLO is written as $CLO_{n}, \quad n=1,2,...N$.
The three dimensional matrix $\mathbf{A}$ is converted into a two dimension matrix $\mathbf{\tilde{A}}$ as
  \begin{equation}\label{vectorA}
  \mathbf{\tilde{A}}=\left[
  \begin{array}{ccccc}
     \mathbf{A}_{1}^{T} & \mathbf{A}_{2}^{T} & . & . & \mathbf{A}_{I}^{T}
   \end{array}
   \right ]
  \end{equation}
  where $\mathbf{A}_{i}^{T}$ is the transpose matrix of $i^{th}$ assessment matrix $\mathbf{A}_{J\times M}$. Matrix $\mathbf{\tilde{A}}$ has the dimension $M\times L$, where $M$ is the total number of students and $L=\sum_{i=1}^{I}J_{i}$ is a total number of questions in all assessments.

CLO to SO mapping matrix is by
\begin{equation}\label{CLOSOmatrix}
\mathbf{CS}=\left[
  \begin{array}{ccccc}
CS_{1,a} & CS_{1,b} & \cdots & CS_{1,k}\\
CS_{2,a} & CS_{2,b} & \ddots & \vdots\\
\vdots & \ddots & \ddots & CS_{N-1,k} \\
CS_{N,a} & \cdots & CS_{N,j} & CS_{N,k}
\end{array}
\right ]
\end{equation}
The matrix element $CS_{n,a}$ is a variable. $CS_{n,a}> 0$ if $CLO_{n}$ maps to SO $a$, otherwise $CS_{n,a}=0$. A non-zero value is a relevance of the CLO to compute the SO. It can take values 1,2,3, for \emph{low, moderate}, and \emph{high} relevance, respectively.\\
Similarly, the CLO to question mapping is given by the following matrix
\begin{equation}\label{CLOquestionMatrix}
\mathbf{CQ}=\left[
  \begin{array}{ccccc}
CQ_{1,1} & CQ_{1,2} & \cdots & CQ_{1,L}\\
CQ_{2,1} & CQ_{2,2} & \ddots &  \vdots\\
\vdots & \ddots & \ddots & CQ_{N-1,L} \\
CQ_{N,1} & \cdots & CQ_{N,L-1} &  CQ_{N,L}
\end{array}
\right ]
\end{equation}
Rows of above matrix contain binary variables $CQ_{n,l}$ that represent the mapping of $n^{th}$ CLO with question $l$, where $l$ maps to $j^{th}$ question of an assessment $i$. $CQ_{n,l}=1$ if $CLO_{n}$ maps to question $l$. \\

%

Student marks in assessment questions are used to compute how well the students have done and what percentage of students have met a certain criteria. Every question contributes to one or more CLOs and every CLO contributes to one or more student outcomes (SO) as shown in \eqref{CLOquestionMatrix} and \eqref{CLOSOmatrix} respectively.
\subsection{CLO Attainment}
This metric is about how well the students have done, in percentage, for each CLO. Attainment of a CLO is derived from the average marks obtained divided by total marks for all questions that maps to the CLO.\\
Let $Q_{i,j}^{tot}$ be the maximum or total marks of question $j$ in assessment $i$. In general, for assessment $i$,
\begin{equation}\label{assessmentMaxMarks}
  \mathbf{Q}_{i}^{tot}=\left[
  \begin{array}{ccccc}
     \mathbf{Q}_{i,1}^{tot} & \mathbf{Q}_{i,2}^{tot} & . & . & \mathbf{Q}_{i,J}^{tot}
   \end{array}
   \right ]
\end{equation}
and
\begin{equation}\label{maxMarks}
  \mathbf{\tilde{Q}}^{tot}=\left[
  \begin{array}{ccccc}
     \mathbf{Q}_{1}^{tot} & \mathbf{Q}_{2}^{tot} & . & . & \mathbf{Q}_{I}^{tot}
   \end{array}
   \right ]_{1\times L}
\end{equation}
Then, the percentage of CLO attainment for $n^{th}$ CLO is given by
\begin{equation}\label{SP_CLO}
  attainmentCLO_{n}[\%]=\frac{\sum_{l=1}^{L}\tilde{B}_{l}*CQ_{n,l}}{\sum_{l=1}^{L}\tilde{Q}_{l}^{tot}*CQ_{n,l}}\times 100
\end{equation}
The operator $\ast$  is used for element-wise multiplication.
\subsection{CLO Weightage Information}
In order to get meaningful results, one should design the CLOs such that there is a uniform distribution of the marks over CLOs in questions to CLO mapping. For example, if a course contains four CLOs then, ideally, each CLO should get $25\%$ weightage.

The ideal case of uniform distribution of marks over the CLOs is seldom realized. In these situations, the CLO weightage information renders a fair picture of $\%$ CLO attainment. The percentage weightage of $n^{th}$ CLO is given by
\begin{equation}\label{CLOweightage}
  WeightageCLO_{n}=\sum_{i}\frac{\sum_{j}^{J_{i}}w(CQ_{n,j})Q_{i,j}^{tot}}{A_{i}^{tot}}\times w(A_{i})
\end{equation}
where $w(CQ_{n,j})$ is the weight of $CLO_{n}$ in question $j$, $w(A_{i})$ is the weight of assessment $i$, and $A_{i}^{tot}$ is the total marks of assessment $i$.
\subsection{Student Achievement per CLO}
Student Achievement per CLO is defined as the percentage of students who are above the expected level as shown in \eqref{passingThreshold}. Expectation or target is a design parameter, one choice of the target could be $\min (B_{i,j},0.7\times Q_{i,j}^{tot})$, i.e., the minimum of the average obtained marks and the $70\%$ of the total marks \cite{reed_hierarchical_2013}.

It counts the number of students who met the criteria by comparing each student marks in a question $j$ of an assessment $i$. If the marks obtained $A_{i,j,m}$ are greater than the passing threshold $PT_{i,j}$, then it increments the counting variable $CPS$ (count pass student) by $1$. Finally, $CPS_{i,j}$ or $CPS_{l}$\footnote{$CPS_{l}$ is a row vector form of $CPS_{i,j}$, similar to \eqref{vectorA} or \eqref{vectorB}} contains number of passed students for each question $j$ in assessment $i$.
Therefore, the average student achievement per CLO is given by
\begin{equation}\label{SA_CLO}
  SA\_CLO_{n}=\frac{\sum_{i=1}^{I}\frac{1}{M_{i}}\sum_{j=1}^{J_{i}}CPS_{i,j}\times CQ_{n,i,j}\times Q_{i,j}^{tot}}{\sum_{i=1}^{I}\sum_{j=1}^{J_{i}}Q_{i,j}^{tot}\times CQ_{n,i,j}}
\end{equation}
where $M_{i}$ is the number of students that participated in the assessment $i$.

\subsection{Student Perception of CLO Attainment}
A course survey is conducted at the end of each semester to gauge students' perception of how well the CLOs were covered in the course. It is the average  of CLO perception from the students.
For each CLO, students provide their input on the scale of $1-5$ where $1$ means CLO is not achieved and $5$ means CLO is achieved completely. Summary of responses is given in following matrix.
\begin{equation}\label{SCmatrix}
\mathbf{SC}=  \begin{bmatrix}
  SC_{1,1} & SC_{1,2} & \cdots & SC_{1,N}  \\
  SC_{2,1} & SC_{2,2} & \cdots & \vdots  \\
  \vdots & \ddots & \ddots & SC_{M-1,N}  \\
  SC_{M,1} & \cdots & SC_{M,N-1} & SC_{M,N} \\
  \end{bmatrix}
\end{equation}
Student's perception of $n^{th}$ CLO attainment is given by
\begin{equation}\label{studentCLO}
  SE\_CLO_{n}=\frac{1}{M}\sum_{m=1}^{M}SC_{m,n}
\end{equation}
\subsection{\emph{x-th} Percentile Marks per CLO}
\emph{x-th} Percentile Marks per CLO is defined as the weighted average of $x-th$ percentile marks divided by total marks of the questions that map to particular CLO.\\
Let $xP_{i,j}$ be the \emph{x-th} percentile marks of question $j$ in assessment $i$. In general, for assessment $i$ we have
\begin{equation}\label{assessmentPercentileMarks}
  \mathbf{xP}_{i}=\left[
  \begin{array}{ccccc}
     \mathbf{xP}_{i,1} & \mathbf{xP}_{i,2} & . & . & \mathbf{xP}_{i,J}
   \end{array}
   \right ]
\end{equation}
and
\begin{equation}\label{percentileMarks}
  \mathbf{\tilde{xP}}=\left[
  \begin{array}{ccccc}
     \mathbf{xP}_{1} & \mathbf{xP}_{2} & . & . & \mathbf{xP}_{I}
   \end{array}
   \right ]_{1\times L}
\end{equation}
Then, the average percentage \emph{x-th} percentile marks per CLO is given by
\begin{equation}\label{xPercentile}
  xPercentileCLO_{n}=\frac{\sum_{l=1}^{L}xP_{l}\ast CQ_{n,l}}{\sum_{l=1}^{L}Q_{l}^{tot}\ast CQ_{n,l}}
\end{equation}
\subsection{SO Attainment}
By using CLO-SO mapping in \eqref{CLOSOmatrix}, course level SO assessment can be achieved. SO attainment for an SO is computed from the CLO attainment of all CLOs that map to the SO. SO attainment is defined as the weighted average of CLOs attainment (in \%).
\begin{equation}\label{weightedAvgSOAttain}
  attainmentSO_{n}=\frac{\sum_{i\in \mathcal{C}^{n}}attainmentCLO_{i}w_{i}}{\sum_{i\in \mathcal{C}^{n}}w_{i}}
\end{equation}
where $\mathcal{C}^{n}$ is the set of CLOs that map to $SO_{n}$ and $w_{i}$ is the weight (or relevance) of $i^{th}$ mapping between CLO and SO.
\subsection{Student Achievement per SO}
It is defined as weighted average of student achievement of CLOs (in \%) that map to a particular SO.
\begin{equation}\label{weightedAvgSASO}
  SA\_SO_{n}=\frac{\sum_{i\in \mathcal{C}^{n}}SA\_CLO_{i}w_{i}}{\sum_{i\in \mathcal{C}^{n}}w_{i}}
\end{equation}
\subsection{Student Perception of SOs Attainment}
Student perception of SOs attainment gives an indirect measurement of SO attainment. This metric is derived from student perception of CLOs attainment in \eqref{studentCLO} and the CLO-SO mapping in \eqref{CLOSOmatrix}.
\begin{equation}\label{weightedAvgSESO}
  SE\_SO_{n}=\frac{\sum_{i\in \mathcal{C}^{n}}SE\_CLO_{i}w_{i}}{\sum_{i\in \mathcal{C}^{n}}w_{i}}
\end{equation}
\subsection{\emph{x-th} Percentile per SO}
\emph{x-th} Percentile per SO uses CLO-Average x-th Percentile \% marks with CLO-SO mapping in \eqref{CLOSOmatrix}.
\begin{equation}\label{weightedAvgxPerSO}
  xPercentileSO_{n}=\frac{\sum_{i\in \mathcal{C}^{n}}xPercentileCLO_{i}w_{i}}{\sum_{i\in \mathcal{C}^{n}}w_{i}}
\end{equation}
\begin{table}[!htbp]
  \centering
  \caption{Evaluation of CLOs for the sample course}
    \begin{tabular}{|c| p{1.3cm}| p{1.3cm}| p {1.3cm}| p {1.3cm}|}
    \toprule
      CLO    & Attainment & Achievement & 50th-Percentile & Student perception\\
    \midrule
    1  & 79.08 & 85    & 82.76 & 85\\
    2  & 82.91 & 91.67 & 87.71 & 72.5 \\
    3  & 78.78 & 78.33 & 82.97 & 72.5\\
    4  & 87.62 & 95    & 89.92 & 67.5\\
    5  & 74.18 & 66    & 76.44 & 77.5\\
    6  & 81.94 & 85    & 79.48 & 77.5\\
    \bottomrule
    \end{tabular}%
  \label{tab:directAssessment}%
\end{table}%

\begin{table*}[t]
\tiny
  \centering
  \caption{Basic Variables to compute evaluations metrics for a sample course. CLOs and questions mapping \eqref{CLOquestionMatrix} is shown in first two rows}
    \begin{tabular}{cccccccccccccccccccc}
    \toprule
          & \multicolumn{4}{c}{midterm}   & Quiz1 & Quiz4 & Quiz3 & \multicolumn{5}{c}{HW1}               & Quiz2 & \multicolumn{5}{c}{Final}             & Class participation \\
    \midrule
    \textbf{CLOs Covered} & \textbf{1,2,3} & \textbf{1} & \textbf{2} & \textbf{3,4} & \textbf{1} & \textbf{6} & \textbf{5} & \textbf{2} & \textbf{2} & \textbf{3} & \textbf{3} & \textbf{3} & \textbf{2} & \textbf{5} & \textbf{5} & \textbf{5} & \textbf{6} & \textbf{6} & \textbf{1-6} \\
    \textbf{Question No.} & \textbf{1} & \textbf{2} & \textbf{3} & \textbf{4} & \textbf{1} & \textbf{1} & \textbf{1} & \textbf{1} & \textbf{2} & \textbf{3} & \textbf{4} & \textbf{5} & \textbf{1} & \textbf{1} & \textbf{2} & \textbf{3} & \textbf{4} & \textbf{5} & \textbf{1} \\
    \textbf{Question Marks} & \textbf{8} & \textbf{8} & \textbf{8} & \textbf{8} & \textbf{4} & \textbf{4} & \textbf{4} & \textbf{0.4} & \textbf{0.4} & \textbf{0.4} & \textbf{0.4} & \textbf{0.4} & \textbf{4} & \textbf{5} & \textbf{10} & \textbf{10} & \textbf{10} & \textbf{10} & \textbf{5} \\
    \hline \\
    Actual Average & 5.22  & 6.6   & 6.99  & 6.69  & 3.25  & 3.664 & 3.12  & 0.4   & 0.4   & 0.4   & 0.352 & 0.128 & 3.68  & 1.9   & 8.3   & 7.2   & 6.1   & 9.3   & 4.7 \\
    Passing Threshold (PT) & 5.22  & 5.6   & 5.6   & 5.6   & 2.8   & 2.8   & 2.8   & 0.28  & 0.28  & 0.28  & 0.28  & 0.128 & 2.8   & 1.9   & 7     & 7     & 6.1   & 7     & 3.5 \\
     No. of Students Above PT& 6     & 9     & 10    & 9     & 9     & 10    & 6     & 10    & 10    & 10    & 8     & 4     & 9     & 7     & 6     & 4     & 4     & 10    & 10 \\
    Minimum Marks & 1.2   & 4.2   & 6     & 4.5   & 1.3   & 3.04  & 2.4   & 0.4   & 0.4   & 0.4   & 0.16  & 0     & 2.4   & 0     & 5     & 5     & 1     & 8     & 4.27 \\
    Maximum Marks & 7.35  & 7.5   & 7.5   & 7.5   & 4     & 3.84  & 3.84  & 0.4   & 0.4   & 0.4   & 0.4   & 0.32  & 4     & 4     & 10    & 10    & 10    & 10    & 4.9 \\
    Standard Deviation & 1.78  & 0.95  & 0.62  & 0.92  & 0.78  & 0.25  & 0.47  & 0.00  & 0.00  & 0.00  & 0.10  & 0.12  & 0.44  & 1.37  & 1.95  & 1.60  & 2.91  & 0.64  & 0.21 \\
    50th Percentile Marks & 5.85  & 6.75  & 7.35  & 6.9   & 3.3   & 3.76  & 3.2   & 0.4   & 0.4   & 0.4   & 0.4   & 0.08  & 3.84  & 2     & 9.5   & 6.5   & 5.5   & 9     & 4.79 \\
    \bottomrule
    \end{tabular}%
  \label{tab:basicVariables}%
\end{table*}%
\begin{table*}[t]
  \centering
  \caption{Mapping of Course Learning Outcomes (CLOs) to Student Outcomes (SOs) \eqref{CLOSOmatrix} as Course Assessment Matrix}\label{courseAssessmentMatrix}
  \begin{tabular}{|c|p {10cm}|c|c|c|c|c|c|c|c|c|c|c|}
\hline
\textbf{CLO} & \textbf{Course Learning Outcomes Description} & \textbf{a} & \textbf{b} & \textbf{c} & \textbf{d} & \textbf{e} & \textbf{f} & \textbf{g} & \textbf{h} & \textbf{i} & \textbf{j} & \textbf{k}\tabularnewline
\hline
\hline
1 & Explain the integrated circuit technology and basic engineering
process steps of CMOS &  &  & 2 &  &  &  &  &  &  &  & \tabularnewline
\hline
2 & Design combinational/sequential logic gates using CMOS
transistor & 2 & 1 &  &  &  &  &  & 1 &  &  & \tabularnewline
\hline
3 & Describe MOS circuit design processes, including CMOS design rules,
symbolic diagrams, and stick diagrams  &  &  & 3 &  &  &  &  &  &  &  & \tabularnewline
\hline
4 & Illustrate the details of layout design and verification, concepts of standard
gate design &  &  &  &  & 2 &  &  &  & 1 &  & \tabularnewline
\hline
5 & Describe the VLSI circuit characterization and perfrommace estimation & 3 & 2 &  &  &  &  &  &  &  &  & 2\tabularnewline
\hline
6 & Analyze and design complex logic gates in standard CMOS technology  & 3 & 3 & 2 &  &  &  &  &  &  &  & \tabularnewline
\hline
\end{tabular}
\end{table*}

\section{Course Level Performance Evaluation}
Direct assessment of an academic program is performed by evaluation of courses in the study plan. If not all, at least a selected subset of the courses is required to find out program's success level. Previous section presented formal formulations of the performance metrics that can be used in course evaluation. This section discusses an implementation of these metrics in evaluation of a sample course. The section starts with setup required for evaluation followed by evaluation results and concludes by discussing issues and concerns.
\subsection{Course Setup for Evaluation}
Course evaluation is computation of performance metrics from basic variable of the course and perform analysis. To compute metrics defined in section II from collected data, each course must have well defined CLOs, CLO to SO mapping as in \eqref{CLOSOmatrix}, CLO to question mapping in each assessment as in \eqref{CLOquestionMatrix}, and passing threshold as defined in \eqref{passingThreshold}.
Table \ref{tab:basicVariables} shows basic variables of a sample course. First two rows show mapping between CLOs and questions for all assessments conducted in the course. Table \ref{courseAssessmentMatrix} shows mapping between CLOs and SOs defined by the course designer. A numeric value in a cell represents a relationship between a CLO and an SO. A value of 1, 2, or 3 indicates that a CLO addresses an SO \emph{slightly, moderately,} or \emph{substantively}. Passing threshold is set to $\min(avg,70\%)$, which is used in computation of student achievement per CLO \eqref{SA_CLO} and student achievement per SO \eqref{weightedAvgSASO}.
\begin{table*}[t]
\small
  \centering
  \caption{Computation of SO attainment from CLO attainment using table \ref{courseAssessmentMatrix} CLO-SO mapping }\label{tab:courseAssessmentMatrixSO}
  \begin{tabular}{|c|c|c|c|c|c|c|c|c|c|c|c|}
\hline
CLO &  a & b & c & d & e & f & g & h & i & j & k\tabularnewline
\hline
\hline
1 &   &  & 79.08 &  &  &  &  &  &  &  & \tabularnewline
\hline
2 &  82.91 & 82.91 &  &  &  &  &  & 82.91 &  &  & \tabularnewline
\hline
3 &  &  & 78.78 &  &  &  &  &  &  &  & \tabularnewline
\hline
4 &   &  &  &  & 87.62 &  &  &  & 87.62 &  & \tabularnewline
\hline
5 &  74.18 & 74.18 &  &  &  &  &  &  &  &  & 74.18\tabularnewline
\hline
6 &  81.94 & 81.94 & 81.94 &  &  &  &  &  &  &  & \tabularnewline
\hline
\hline
  SO attainment  & 79.68 & 79.68 & 79.94 &  & 87.62 &  &  & 82.91 & 87.62 &  & 74.18 \tabularnewline
\hline
  Weighted SO attainment  & 79.27 & 79.52 & 79.77 &  & 87.62 &  &  & 82.91 & 87.62 &  & 74.18 \tabularnewline
\hline
 Relevance  & 3 & 2 & 3 &  & 2 &  &  & 1 & 1 &  & 2\tabularnewline
\hline
\end{tabular}
\end{table*}
\subsection{Performance Evaluation}
This section presents computed values of metrics defined in section II for the sample course. Table \ref{tab:directAssessment} presents a summary of CLO related metrics and table \ref{tab:courseAssessmentMatrixSO} shows how to compute SO attainments from CLO attainments.
\subsubsection{CLO Attainment}
\begin{figure}[!htbp]
  \centering
  \includegraphics[width=3.5in]{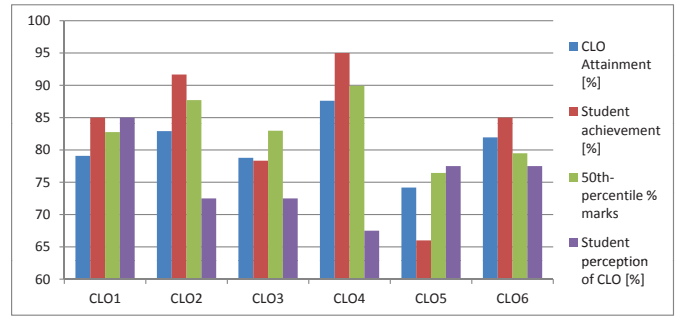}\\
  \caption{CLO Performance Evaluation}\label{CLOs}
\end{figure}

CLO attainment for the sample single section course is shown in Fig. \ref{CLOs}. CLO attainment quantifies the student attainment level of particular CLO through the percentage marks allocated to that CLO. Since this is a percentage value of average marks obtained in the questions maps to a particular CLO, therefore, it is necessary to either distribute the marks uniformly over the CLOs or give an explicit evidence of CLO to marks ratio.
\subsubsection{CLO Weightage}
The CLO weightage for a sample single section course is shown in Fig. \ref{CLOweightage}. The CLO attainment and student achievement of CLO are based on these weightages.
\begin{figure}[H]
  \centering
  \includegraphics[width=2.5in]{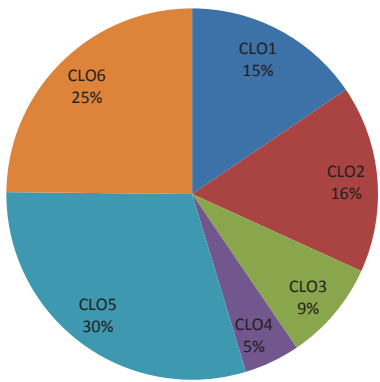}\\
  \caption{CLOs Weightage}\label{CLOweightage}
\end{figure}
\subsubsection{Student Achievement per CLO}
Student achievement per CLO for the sample single section course is shown in Fig. \ref{CLOs}. It is the percentage number of students that meet or exceed the target or expectations. There is an upper limit for the target ($70\%$) but there is no lower limit and it depends upon the average marks. We can get absolute student achievement by fixing the target, for example, with target value of $60\%$.

\subsubsection{\emph{50-th} Percentile per CLO}
The \emph{50-th } percentile for the sample course is shown in Fig. \ref{CLOs}. It shows the percentage median marks for each CLO.

\subsubsection{Student Perception of CLOs Attainment}
 For each CLO, student perception of CLO attainment can be on the scale of \textbf{1} to \textbf{5}, 1 mean \emph{strongly disagree} to 5 for \emph{strongly agree}. Student perception of CLO attainment for the sample course is shown in Fig. \ref{CLOs}. There were 10 students but 8 participated in the course survey.

\subsubsection{SO Attainment}
\begin{figure}[!htbp]
  \centering
  \includegraphics[width=3.5in]{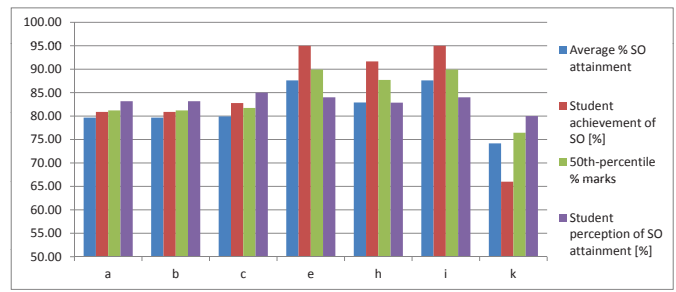}\\
  \caption{SO Performance Evaluation}\label{SOs}
\end{figure}

Bar graphs for SO attainment are shown in Fig. \ref{SOs}. These levels are averages of CLO attainments that map to particular SO, therefore, the health of CLO attainments and CLO-SO mapping is critical.

\subsubsection{Student Achievement per SO}
Student achievement for the sample course is given 
 in Fig. \ref{SOs}. This is a derived value from CLO achievements and depicts the percentage number of students achieved the set target averaged over the CLOs mapped to that SO.
\subsubsection{\emph{50-th} Percentile per SO}
The \emph{50-th} percentile per SO values are derived from \emph{50-th} percentile per CLO. Fig. \ref{SOs} depicts the \emph{50-th} percentile or median marks for each mapped SO.

\subsubsection{Student Perception of SOs Attainment}
The student perception of SOs attainment is shown in Fig. \ref{SOs}. It is an indirect measurement obtained from the course exit survey where students provide their feedback about the CLOs attainment.
\subsection{Issues and Guidelines}
Course designer is responsible to establish quality mapping between CLOs and SOs. Course instructor is responsible for CLOs to questions mapping for all assessments. Quality of these mappings have direct impact on the evaluation results as discussed in rest of this section.
\subsubsection{Relationship of Questions, Marks distribution and CLOs}
It has been observed that questions to CLO mapping requires uniform marks distribution over the CLOs. The quantitative measurement of CLOs provides the baseline data for direct assessment, therefore, questions to CLOs mapping is critical in direct assessment. CLOs should be designed in such a way that they cover all the core topics (qualitative equality) and course assessments should cover all CLOs with uniform marks distribution over the CLOs (quantitative equality). Similar measures are required in capstone project rubrics' design. Capstone project is an important entity of program in which students apply the knowledge gained during the course of the program to solve the engineering problems. The capstone project rubrics map to CLOs and these CLOs usually cover all the SOs. Since the sample size in this assessment is not as large as of direct assessment therefore results may differ in these assessments.

\subsubsection{Questions to CLOs Mapping Approaches}
Due to the many-to-many mapping between questions and CLOs, a common question arises about the weights of a question that maps to multiple CLOs. There are three possibilities:
\begin{itemize}
  \item One-to-many mapping with equal weights
  \item One-to-many mapping with proportional weights
  \item One-to-one mapping between questions and CLOs.
 \end{itemize}
  In this manuscript, equal weights have been used in questions to CLOs mapping. The proportional weights add one more level of complexity for the faculty and hence more chances of errors. One-to-one mapping is another attractive solution which eliminates the weight problem because in this case one question can be mapped to one CLO at most. In this technique many questions can be mapped to one CLO but converse is not possible. Proportional weights and one-to-one schemes require a proper design of CLOs and the mapping table between questions and CLOs.

\subsubsection{CLOs to SOs Mapping within a Course}
There are three choices:
\begin{itemize}
  \item One CLO can be mapped to any number of SOs without weights (one-to-many mapping without weights)
  \item One CLO can be mapped to any number of SOs with weights (one-to-many mapping with weights)
  \item One CLO can be mapped to one SO only (one-to-one mapping) \cite{dargham_linking_2012}
\end{itemize}
In this manuscript, one-to-many mapping with weights has been used as shown in Table \ref{courseAssessmentMatrix}. One-to-many mapping without weights assumes equal weights across all SOs mapped to a particular CLO. A straight forward way of mapping is one-to-one mapping which does not require weights but again the design of CLOs is important in this case.

\section{Program Level Performance Evaluation}
Achievement of student outcomes (SO) and program educational objectives (PEO) are the basis for program assessment. PEOs are computed from prgram level SOs and indirect methods of assessment such as graduate exit surveys, alumni surveys, and  employer surveys \cite{pierrakos_comprehensive_2013}.

To measure the achievement level of SOs, direct as well as indirect methods are used as shown in \ref{AssessmentProcess}.
Previous section presents direct assessment approaches and results of SO achievements derived from students' performance in courses. These course level SOs achievements are used to build the program level direct assessment of SOs. Direct method uses a subset of courses in the study plan to compute program level SOs.

\begin{table}[!htbp]
  \centering
  \caption{Relationship between the Program Educational Objectives and Student Outcomes}
    \begin{tabular}{cccccccccccc}
    \toprule
    \multirow{2}[4]{*}{\textbf{PEOs}} & \multicolumn{11}{c}{\textbf{Student Outcomes}} \\
 \cline{2-12}
          & a     & b     & c     & d     & e     & f     & g     & h     & i     & j     & k \\
          \cline{2-12}
    I     & x     & x     & x     & x     & x     &       &       &       &       &       & x \\
    II    & x     & x     & x     &       & x     &       &       & x     &       &       & x \\
    III   &       &       &       & x     &       &       & x     &       &       &       &  \\
    IV    &       &       &       &       &       & x     &       & x     & x     & x     &  \\
    \bottomrule
    \end{tabular}%
  \label{tab:PEOSOmapping}%
\end{table}%

\begin{table*}[htpb]
\centering
\caption{Program Assessment Matrix based on Courses (Substantial: 3, Moderate: 2, Slight: 1, None: blank)}\label{programAssessmentMatrixCourses}
\begin{tabular}{|c|p {5cm}|c|c|c|c|c|c|c|c|c|c|c|}
\hline
\multicolumn{2}{|c|}{ Assessment} & a & b & c & d & e & f & g & h & i & j & k\tabularnewline
\hline
\hline
\multirow{17}{*}{\begin{turn}{90}DIRECT and INDIRECT\end{turn}}
& English &  &  &  &  &  &  & 3 &  & 2 &  & \tabularnewline
\cline{2-13}
 & Calculus I,II,III & 3 &  &  &  &  &  &  &  &  &  &2 \tabularnewline
 \cline{2-13}
 & Linear Algebra & 3 &  &  &  &  &  &  &  &  &  & 2\tabularnewline
 \cline{2-13}
 & Differential Equations & 3 &  &  &  &  &  &  &  &  &  & 2\tabularnewline
\cline{2-13}
 & Probability and Stats & 3 &  &  &  &  &  &  &  &  &  & \tabularnewline
\cline{2-13}
 & Discrete Math & 3 &  &  &  &  &  &  &  &  &  & \tabularnewline
\cline{2-13}
 & Physics I,II & 3 & 3 &  &  &  &  &  &  &  &  & \tabularnewline
 \cline{2-13}
 & Computer Programming I, II & 3 & 3 & 2 &  &  &  &  &  & 2 & 2 & \tabularnewline
 \cline{2-13}
 & Data Structure &  & 2 &  &  &  &  &  &  &  & 3 & \tabularnewline
 \cline{2-13}
 & Operating Systems & 3 & 2 &  &  &  &  &  &  & 2 & 1 & \tabularnewline
 \cline{2-13}
 & Fundamental of Database &  & 3 & 2 &  &  &  &  &  & 2 &  & \tabularnewline
 \cline{2-13}
 & Software Engineering &  & 3 & 2 &  &  &  &  &  & 2 & 2 & \tabularnewline
\cline{2-13}
 & Digital Logic Design & 3 & 3 & 1 &  & 1 &  &  &  &  &  & \tabularnewline
\cline{2-13}
 & Electric Circuits & 3 & 3 & 2 &  & 1 &  &  &  &  &  &3 \tabularnewline
\cline{2-13}
 & Electronics & 3 & 3 &  &  &  &  &  &  &  &  & \tabularnewline
\cline{2-13}
 & Computer Architecture & 2 & 3 & 1 &  &  &  &  &  & 1 & 1 & \tabularnewline
 \cline{2-13}
 & Computer Organization and Design & 1 & 2 & 1 &  &  &  &  & 1 &  & 1 & \tabularnewline
\cline{2-13}
 & Microprocessor & 3 & 2 & 3 &  &  &  &  &  &  &  & 2 \tabularnewline
 \cline{2-13}
 & Professional Ethics &  &  & 2 &  &  & 3 &  &  &  &  & \tabularnewline
\cline{2-13}
 & Embedded Systems & 2 & 2 &  &  & 1 &  &  &  &  &  &1 \tabularnewline
\cline{2-13}
 & Signals and Systems & 3 & 3 & 3 &  & 1 &  &  &  &  &  & \tabularnewline
\cline{2-13}
 & Communication Systems & 3 & 1 & 2 &  &  &  &  &  &  &  & \tabularnewline
\cline{2-13}
 & Digital Signal Processing & 2 & 3 & 2 &  &  &  &  &  &  &  & 1 \tabularnewline
\cline{2-13}
 & Control Engineering & 3 & 3 & 1 &  & 1 &  &  &  &  &  & \tabularnewline
\cline{2-13}
 & VLSI Design & 3 & 2 & 3 &  & 2 &  &  & 1 & 1 &  &2 \tabularnewline
\cline{2-13}
 & Capstone Design Project & 2 & 3 & 3 & 3 & 3 & 3 & 3 & 3 & 3 & 3 & 3\tabularnewline
\hline
\end{tabular}
\end{table*}

\begin{table*}[htpb]
\centering
\caption{Program Assessment Matrix based on Surveys (Substantial: 3, Moderate: 2, Slight: 1, None: blank)}\label{programAssessmentMatrixSurveys}
\begin{tabular}{|c|p {5cm}|c|c|c|c|c|c|c|c|c|c|c|}
\hline
\multicolumn{2}{|c|}{ Assessment} & a & b & c & d & e & f & g & h & i & j & k\tabularnewline
\hline
\hline
\multirow{8}{*}{\begin{turn}{90}INDIRECT\end{turn}}
& Graduate Exit Survey & 2 & 2 & 2 & 2 & 2 & 2 & 2 & 2 & 2 & 2 & 2\tabularnewline
\cline{2-13}
 & Student Course Exit Survey & 2 & 2 & 2 & 2 & 2 & 2 & 2 & 2 & 2 & 2 &2 \tabularnewline
 \cline{2-13}
& Capstone Project Rubrics  & 2 & 2 & 2 & 2 & 2 & 2 & 2 & 2 & 2 & 2 &2 \tabularnewline
\cline{2-13}
 & Course Assessment by Instructor(s)& 2 & 2 & 2 & 2 & 2 & 2 & 2 & 2 & 2 & 2 & 2\tabularnewline
\cline{2-13}
 & Course Assessment by Reviewer(s) & 3 & 3 & 3 & 2 & 2 & 2 & 2 & 2 & 2 & 3 & 3\tabularnewline
\cline{2-13}\\
\cline{2-13}
& Alumni Survey & 1 & 1 & 1 & 1 & 2 & 2 & 2 & 2 & 2 & 3 & 3\tabularnewline
\cline{2-13}
& Employer Survey & 1 & 1 & 1 & 1 & 2 & 2 & 2 & 2 & 2 & 3 & 3\tabularnewline
\cline{2-13}
& Advisory Board Feedback & 1 & 1 & 1 & 1 & 2 & 2 & 2 & 2 & 2 & 3 & 3\tabularnewline
\hline
\end{tabular}
\end{table*}

\subsection{Program Setup}

The program educational objectives are well-defined and published objectives of an institution.
Department of Computer Engineering at Taif University has published the following program educational objectives.\\
Within two to three years of graduation, our Computer Engineering Graduates will:
\begin{description}
  \item[(I)] Be employed as computer engineering professionals in public or private sector and/or perform at a satisfactory level in graduate programs.
  \item[(II)] Possess the necessary knowledge in fundamental theories, techniques and tools to solve computer engineering problems, design hardware/software systems and improve their performance.
  \item[(III)] Demonstrate communication and inter-disciplinary collaboration skills, and leadership qualities.
  \item[(IV)] Engage in lifelong learning activities within a professional, legal and ethical framework.
  \end{description}

A mapping between PEOs and SOs is required to derive PEO achievment values from program level SO values. This mapping should be designed by stakeholders responsible for program assessment. Table \ref{tab:PEOSOmapping} present a sample mapping between PEOs and SOs, which is used in this paper.

 Student outcomes are the defined skills that a student should have; after successfully completing the educational program. An educational program consists of \emph{general knowledge}, \emph{foundation core}, \emph{advanced core}, and \emph{elective} courses. All courses in a program can be used to derive SO achievement levels or a set of witness courses can be selected to ensure that all SO are properly covered. Required advanced courses in the major field of study are obvious candidates for this set. Courses such as mathematics, physics, chemistry, and English might be included as long as they consistently address outcomes. Elective courses or courses whose content varies from one offering to another (so that the outcomes might not be addressed in a particular offering) should not be included \cite{felder_designing_2003}.

Core courses of a program include courses related to the specific program along with math and basic sciences courses. All core courses must have mapping between CLOs and SOs \cite{felder_designing_2003}. Electives courses for which CLOs are not consistent in subsequent offering, should not be considered in program assessment.

A program assessment matrix (PAM) can be constructed as shown in Table \ref{programAssessmentMatrixCourses} for direct assessment based on students' performance in the courses. For courses to SOs mapping, each course gives a relevance number ($1-3$) for each mapped SO in the Table \ref{programAssessmentMatrixCourses}. This is calculated from the weights along each SO column in Table \ref{courseAssessmentMatrix}. It has been observed that program assessment Table \ref{programAssessmentMatrixCourses} should contain the weights for credit hours ($1-4$) along the SOs columns, such that weight is one for one credit hour, two for two credit hours and so on. It will provide more fair values of attainments, achievements, and percentiles.

An identical matrix should also be used for indirect assessment based on course exit survey that shows students' perception of SO achievement. Surveys are indirect assessment methods and a PAM can be created as shown in Table \ref{programAssessmentMatrixSurveys}. First five rows of the table are related to the course offerings in a program while last three row are about perception from the stakeholders in the marketplace about the program itself. Entries of 1, 2, and 3 for each SO in the matrix denote \emph{slight, moderate}, and \emph{substantial} relevance to the program outcomes \cite{ezeldin_international_2013}.
These matrices provide a brief summary of how SOs are assessed when attempting to raise the attainment level of a particular SO.

Program level direct assessments are obtained by putting the weighted average of SO attainment values from all course level SO attainment tables in program assessment Table \ref{programAssessmentMatrixCourses}. For example, the weighted average of SO attainment (second last row) from Table \ref{tab:courseAssessmentMatrixSO} in the corresponding course row of Table \ref{programAssessmentMatrixCourses}. Similar matrix can be used to compute student perception of SOs for each course as indirect assessment. Since capstone project contains both direct and indirect assessment of CLOs and SOs, SO attainment can be directly used here.


\subsection{Performance Evaluation}
There are multiple sources of data to compute the attainment of SOs at program level. Bulk of the data and evidences come from courses offered during the program and students' performance in the courses. For every course students provide their perception on attainment of SOs using course survey. Assessment of capstone graduation project is done directly based on students' performance as well as done indirectly based on rubrics.

\begin{figure}[!htbp]
  \centering
  \includegraphics[width=3in]{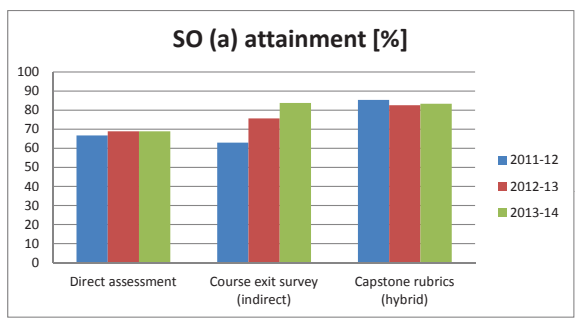}\\
  \caption{Comparison of SO \textbf{''a''} attainment}\label{programSOattainment}
\end{figure}

Table \ref{tab:programSOattaiment} shows attainment trend of SO \textbf{''a''} from different direct and indirect assessment methods and same data is shown in Fig. \ref{programSOattainment}. The intra-assessment trend for direct assessment shows marginal improvement in three years so as in the case of capstone project rubrics. But there is significant increase in students' perception of SO attainment. This all due to the increasing standards of education quality during the ABET preparation and after the ABET accreditation. The inter-assessments comparison shows that direct assessment levels are constant despite the monotonic increase in students' perception. It exhibits the gradual increase of quality standards, realized by students. The capstone project rubrics assessment have highest values among all assessments which are always anticipated because of small sample sizes and for the graduating students. Attainment of other SOs can be computed and presented similarly.

\begin{table}[!htbp]
  \centering
  \caption{Comparison of SO \textbf{''a''} attainment }
    \begin{tabular}{rrrr}
    \toprule
    \textbf{Assessment type} & \textbf{2011-12} & \textbf{2012-13} & \textbf{2013-14}\\
    \midrule
    Direct assessment & 66.75    & 68.9 & 68.89 \\
    Course exit survey (indirect) & 62.98    & 75.67 & 83.77 \\
    Capstone rubrics (hybrid) & 85.4    & 82.58 & 83.36 \\
    \bottomrule
    \end{tabular}%
  \label{tab:programSOattaiment}%
\end{table}%

\begin{table}[!htbp]
  \centering
  \caption{Attainment of Program Educational Objective I }
    \begin{tabular}{rrrr}
    \toprule
    \textbf{Assessment type} & \textbf{2011-12} & \textbf{2012-13} & \textbf{2013-14} \\
    \midrule
    Direct assessment & 70    & 68.5 & 67 \\
    Graduate exit survey & 76    & 80 & 83 \\
    Alumni survey & 89    & 86 & 87 \\
    Employer survey & 79    & 85 & 84 \\
    \bottomrule
    \end{tabular}%
  \label{tab:PEOattainment}%
\end{table}%

\begin{figure}[!htbp]
  \centering
  \includegraphics[width=3in]{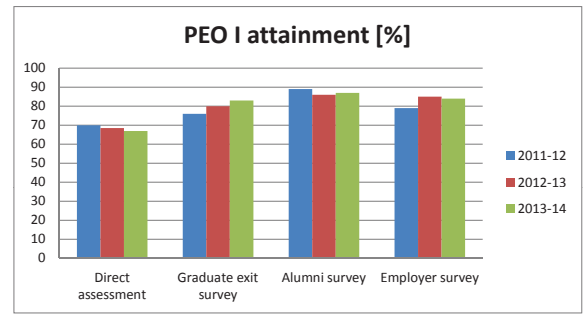}\\
  \caption{Attainment of Program Educational Objective I }\label{PEOattainment}
\end{figure}
The trend of PEOs attainment can be represented for direct assessment, graduate exit survey, alumni survey, and employer survey. As an example Table \ref{tab:PEOattainment} shows attainment of \textbf{PEO I} and a graph representation is shown in Fig. \ref{PEOattainment}. Similar process can be used to compute and present attainment of PEO II, III, and IV for the program.

\section{Analysis and Interpretations}
This section provides a detail analysis of course and program level evaluations based on the formulated evaluation metrics. The implementation of these metrics have revealed several new directions and interpretations.

\subsection{Analysis and Interpretations: Course level}
\begin{figure}[!htbp]
  \centering
  \includegraphics[width=3in]{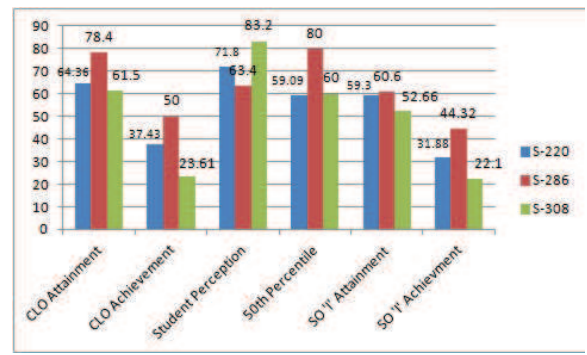}\\
  \caption{Comparison of computed metrics for three sections of the same course for a CLO-SO mapping}\label{CourseMetrics01}
\end{figure}

\begin{figure}[!htbp]
  \centering
  \includegraphics[width=3in]{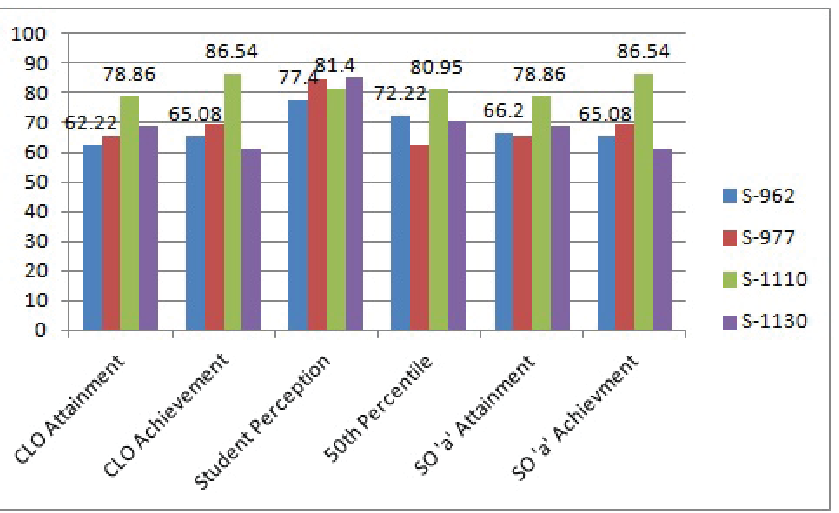}\\
  \caption{Comparison of computed metrics for four sections of the same course for a CLO-SO mapping}\label{CourseMetrics02}
\end{figure}

\begin{figure}[!htbp]
  \centering
  \includegraphics[width=3in]{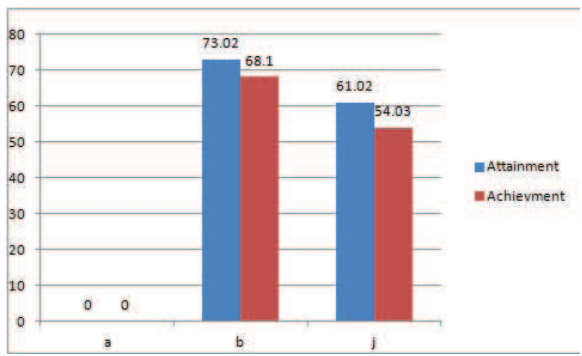}\\
  \caption{Relationship of SO attainment and achievment and coverage of SO in the course}\label{CourseMetrics04}
\end{figure}

\begin{figure}[!htbp]
  \centering
  \includegraphics[width=3in]{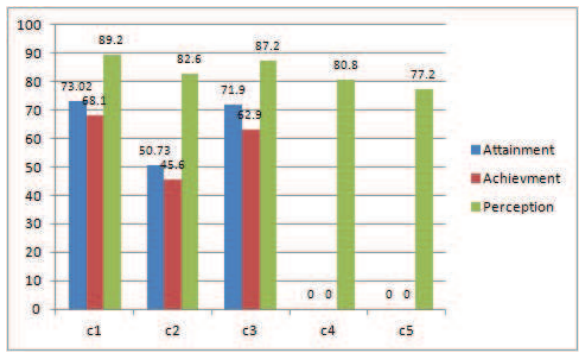}\\
  \caption{CLO metrics and coverage of CLO in the course}\label{CourseMetrics05}
\end{figure}

Course evaluation produces quantitative values of attainment, achievement, and $x-th$ percentile metrics for CLOs and SOs. For a course, relative values of these metrics provide insight into what happened in the course and zero value for a metric indicates that topics related to the corresponding CLO or SO are either not covered in the course or data was not collected for evaluation. For a multi-section course, these metrics can point to lack of coordination among course instructors, and difference in teaching and evaluation standards,

If a course instructor does not cover some CLOs then corresponding metrics values will be zero as shown in figure \ref{CourseMetrics04} and \ref{CourseMetrics05}. For the sample course, CLO 4 and 5 are not covered and since these two CLOs maps to SO ''a'', so the SO is also not covered in the course. For multi-section courses, a zero value for any of the defined metrics in some of the section indicates lack of coordination among course instructors. Figures \ref{CourseMetrics01} and \ref{CourseMetrics02} points to different teaching and evaluation criteria among different instructors of the same course.

\begin{figure}
  \centering
  \includegraphics[width=3in]{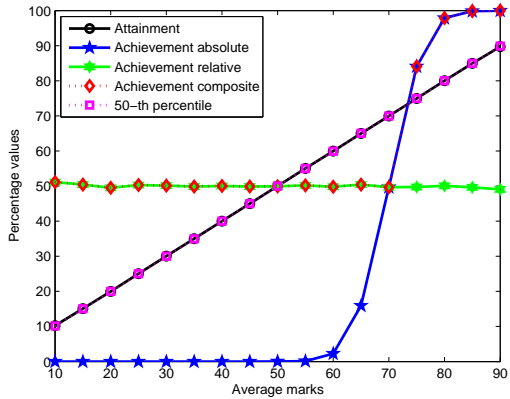}\\
  \caption{Comparison of CLO attainment, student achievement, and \emph{50-th} percentile}\label{CLOattenAchPer_trunc}
\end{figure}

In order to explain the relationship, CLO attainment, student achievement, and \emph{50-th} have been plotted against the students' average marks in Fig. \ref{CLOattenAchPer_trunc}. These graphs show the three evaluation metrics' values for a range of average marks associated with a particular CLO. In this figure, the average marks are obtained from normal distribution  mean (average) with standard deviation $5$. Number of students is $30$ and the results are averaged over $1000$ iterations. From this figure, it can be seen that CLO attainment is a linear function of questions' average marks mapped to that CLO. The attainment is equal to the \emph{50-th} percentile because the mean and median of normal distribution are equal. The composite student achievement remains almost constant up to $70\%$ average marks due to the passing threshold $\min(averageMarks,70\%TotalMarks)$. For normally distributed marks, there are always $50\%$ students below the average value and $50\%$ students are above the average value, hence, student achievement remains constant at $~50\%$ value. When the average marks go above $70\%$, the passing threshold shifts from average value to $70\%$ and all the students with marks greater than $70\%$ contribute to the student achievement. At about $80\%$ average marks, the student achievement reaches to $100\%$ value because all the students even with the standard deviation of $5$ now lie above $70\%$ threshold. The relative achievement always remains around $50\%$. The absolute achievement for $\alpha =0.7$ crosses the $50\%$ value at average marks equal to $70$.

\begin{figure}[!htbp]
  \centering
  \includegraphics[width=3in]{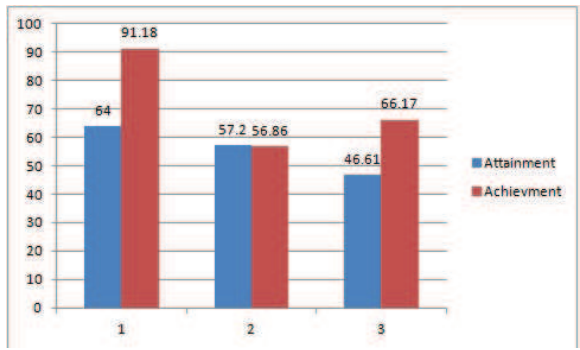}\\
  \caption{Relationship of SO attainment and achievment}\label{CourseMetrics03}
\end{figure}

The designed evaluation metrics give comprehensive results when considered collectively as shown in figures \ref{CourseMetrics01} to \ref{CourseMetrics05}.
\begin{enumerate}
  \item \emph{Attainment and Achievement}:
   The relationship between attainment and student achievement (absolute, composite) is linear for average marks greater than set target, i.e., for sufficiently large population size and normal distribution of obtained marks, high values of attainment corresponds to high values of achievements and the low attainment expects low achievement. Attainment and achievement are independent for achievement level below the threshold. If the distribution is not normal then linearity is not guaranteed. For example, if there are $10$ students and $9$ students secure $50$ marks (out of $100$) and one student get $10$ marks then the composite achievement is $90\%$ but the attainment is $46\%$
  \item \emph{Attainment and Percentile}:
   The \emph{50-th} percentile (or median) gives an additional information about the health of attainment. It is also called location parameter. Median close to attainment indicates normal distribution of marks.
  \item \emph{Achievement and Percentile}:
   If \emph{50-th} percentile (median) is equal to the target value of achievement then the achievement is equal to $50\%$. Median values above the the achievement target shows that more students have met the expectation and value of achievement will be high. Median values less than the achievement target results in the achievement level less than $50\%$.
  \item \emph{Attainment, Achievement, and Percentile}:
   Attainment and \emph{50-th} percentile (median) have the same units, i.e., average and median marks in a particular CLO, whereas, student achievement gives the number of students. If attainment and median are similar (normal distribution) and have high values then the absolute and composite achievements will also be high because more number of students would have marks greater than the set target, whereas, the relative achievement will remain flat at $50\%$ because of normally distributed marks. Conversely, if attainment and median have low levels then the composite achievement becomes $~50\%$. Note that the absolute achievement is proportional to the attainment and median near the target and becomes independent for the average marks sufficiently less or greater than the target value.
\end{enumerate}

\begin{table}[!htbp]
  \centering
  \caption{Interpretations of attainment, student achievement, and median values. Other combinations are not common}
    \begin{tabular}{p{1cm} p{1cm} p {1cm} p {3cm}}
    \toprule
      Attainment & Student Achievement & 50th-Percentile (Median) & Interpretations \\
    \midrule
    High & High & High & If values are greater than 80, then course and assessments should be reviewed \\
    Low  & High & Low  & Majority have low grades and few have worse; special case of one student with low grade\\
    Low  & Low  & Low  & Majority have low grades and few have good\\
    \bottomrule
    \end{tabular}%
  \label{tab:interpretation}%
\end{table}%

\subsection{Analysis and Interpretations: Program level}
Program assessment depends mainly on the course level direct assessments. Other sources of program assessments are indirect assessments through surveys which complement the results obtained from direct assessments.

SOs and PEOs level obtained from various sources are finally analyzed through statistical tool. Results from two datasets are compared with t-test whereas, more than two datasets are compared with analysis of variance (ANOVA) test and graphical plots \cite{montgomery2009engineering}.

The indirect assessments of SOs and PEOs are very subjective. One cannot compare the SO's and PEO's levels in direct and indirect assessment using the same weights. The indirect assessments, especially, the surveys, require extra efforts in determining the the weights relative to the direct assessment. These weights depend upon the type of employer, quality of advisory board, and the interaction level with alumni. In general, these weights may differ institute by institute and region by region.

\section{Conclusions}
Established universities have assessment and evaluation processes in-place but for less established and newer institutions learning, understanding, and establishing these processes are big challenges. This paper describes a complete evaluation process of assessment data for an educational degree program with the help of various evaluation metrics and quantification approaches of PEOs and SOs attainment levels. The main focus of this paper is the formulation and implementation of performance metrics for CLOs, SOs, and PEOs evaluation. Analysis and interpretation of processed data along with the relationship among questions, CLOs and SOs is presented to give an insight into evaluation process.

\bibliographystyle{ieeetr}
\bibliography{assessbib}
\end{document}